# Membraneless flow battery leveraging flow-through heterogeneous porous media for improved power density and reduced crossover

M.E. Suss,[a] K. Conforti,[a] L. Gilson,[b] C.R. Buie,[b] and M.Z. Bazant,[a,c,*]

[a] Department of Chemical Engineering, MIT, Cambridge, MA, USA
[b] Department of Mechanical Engineering, MIT, Cambridge, MA, USA
[c] Department of Mathematics, MIT, Cambridge, MA, USA
* Corresponding author, Email: bazant@mit.edu

**Abstract**

A key factor preventing the market penetration of renewable, intermittent energy sources, such as solar, wind and wave, is the lack of cost-effective energy storage options to counteract intermittency. Here, we propose and demonstrate a novel flow battery architecture that replaces traditional ion-exchange membranes with less expensive heterogeneous flow-through porous media. We present an experimentally-validated model which demonstrates that our architecture promises reduced crossover of reactive species compared to typical membraneless systems employing co-laminar flows in open channels. In addition, our prototype battery exhibits significantly improved power density (0.925 W/cm$^2$) and maximum current density (3 A/cm$^2$) compared to previous membraneless systems.

**Introduction**

Redox flow batteries have the potential to provide geographically flexible and highly efficient energy storage,[1–3] but their market penetration has been significantly limited by high system costs. In conventional redox flow batteries, the single most expensive component is typically the ion-exchange membrane that separates the anode and cathode.[4–6] It is widely believed that such membranes with highly charged pore-surface groups are required to prevent the crossover of similarly charged active species through electrostatic repulsion. Besides their high direct costs, ion-exchange membranes also contribute indirectly to the system cost via significant resistive losses and solvent management issues due to electro-osmotic drag.[2,3,5]

Laminar-flow or "membraneless" cells circumvent these issues by simply removing the membrane and allowing the free diffusion of active species between laminar streams co-flowing through the anode and cathode compartments.[7,8] Such a cell was first demonstrated in 2002 using vanadium chemistry,[9] and a variety of other redox couples have since been considered.[10–12] However, the vast majority of membraneless cells only operate as fuel cells, delivering energy but not regenerating reactant streams.[7] Flow batteries for energy storage, by contrast, operate in closed-loop cycles of discharge (galvanic) and charge (electrolytic) modes to regenerate reactant streams of fixed total volume. To our knowledge, the only prior report of closed-loop cycling operation of a membraneless cell demonstrated a single cycle at 20% round-trip energy efficiency (using vanadium redox chemistry).[13] The overwhelming majority of membraneless systems utilize co-flowing stream in open channels to separate reactants,[7,11,13,14] yet the results to date show that this cell architecture results in high crossover rates between the co-flowing streams, preventing effective battery cycling.[11,13–16] The work by Hollinger et al. and Da Mota et al. used instead a nanoporous separator between co-flowing streams in open channels in an effort to reduce crossover, but only demonstrated single pass performance (fuel cell mode) and not closed-loop cycles (flow battery mode).[10,17] Innovations in membraneless cell architecture to minimize reactant crossover and deliver high performance are required to unlock their potential as low cost grid-scale energy storage systems.

Here, we present a novel membraneless flow battery architecture designed to minimize reactant crossover and enable exceptional power density, and demonstrate this architecture using $H_2$-$Br_2$ chemistry (Figure 1a). The key novelty is our combination of the concepts of flow-through porous electrodes with flow perpendicular to electric field (for high current and power densities) and nanoporous separators (for reduced crossover) to enable a high performance, cyclable membraneless flow battery. While previous membraneless cells have used flow-through porous electrodes (albeit with flow largely parallel to electric field),[13,18,19] or nanoporous separators,[10,17] no previous system to our knowledge has combined these two concepts. This combination results in a flow-through, heterogeneous pore structure cathode assembly, consisting of an electrode material with high porosity and > 10 µm diameter pores for efficient fluid flow, and a thin layer with smaller < 1 µm diameter pores and lower porosity. This small-pore layer, which we term a "dispersion-blocker", is not an ion exchange membrane, as it does not contain highly charged surface groups that electrostatically block similarly charged ions. We show that the dispersion blocker promises to limit reactant crossover via eliminating hydrodynamic dispersion (fast convective mixing) and minimizing molecular diffusion between liquid streams, that our novel architecture allows for the highest maximum power density (0.925 W/cm$^2$) and current density (3 A/cm$^2$) reported for a membraneless cell, and demonstrate proof-of-concept closed-loop cycling.

**Theory and calculations**

In order to study crossover in our cell architecture, we developed a 2D numerical model of our experimental cell (shown schematically in Figure 1a). The model assumes laminar co-flowing streams with a bromine/hydrobromic acid solution flowing through a porous cathode (the oxidant) and hydrobromic acid solution flowing through an open channel (the electrolyte). Between these two flows, a thin dispersion blocker layer is included, within which the fluid is assumed quiescent. Our model captures the diffusive transport of solvated liquid bromine ($Br_2$), tribromide ($Br_3^-$) and bromide ions ($Br^-$) coupled to the (fast) complexation reaction involving these species ($Br_2 + Br^- \rightarrow Br_3^-$, $K$ = 16.7).[5,20] Previous transport models of $H_2$-$Br_2$ energy systems neglected bromine/bromide complexation, despite an equilibrium constant much greater than unity.[5,16,21] As we are primarily interested in crossover, the model neglects the secondary effects of electromigration and electrode reactions. The transport equations for $Br_2$, $Br^-$, and $Br_3^-$ can be combined by employing the molar conservation of elemental bromine (Br), electroneutrality, and complexation equilibrium to arrive at a single governing transport equation:

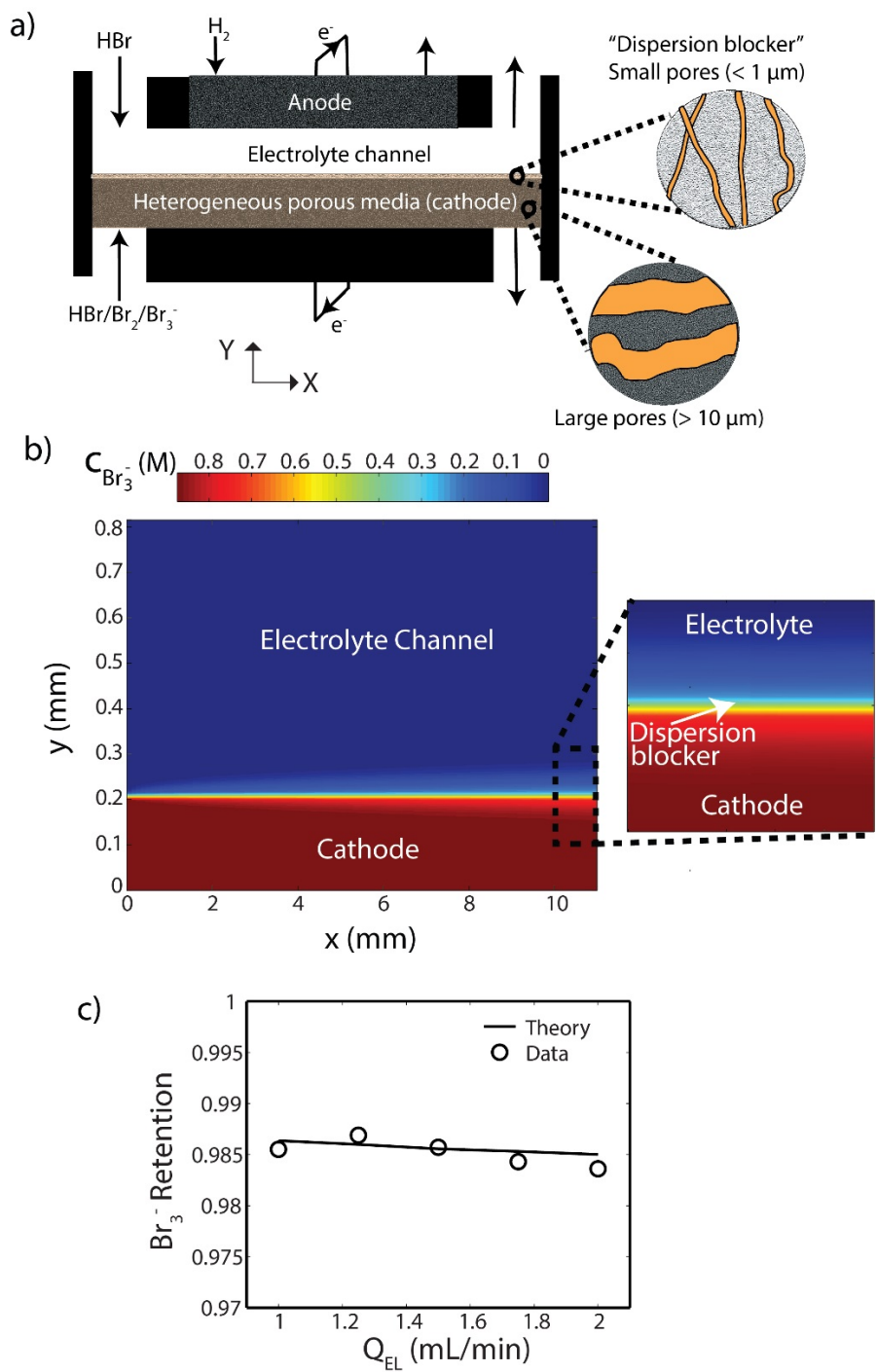

**Figure 1**: a) Schematic of the novel flow battery architecture proposed and demonstrated in this work. b) Example solution of Eq. 1 for an area-average velocity in the electrolyte channel of 2.8 cm/s. c) Tribromide retention predicted by our model and that measured from the experimental cell, versus the electrolyte flowrate, $Q_{EL}$.

$$\frac{\partial}{\partial x}\left[U\left(\frac{2c_{Br_3^-}}{K(c_{H^+}-c_{Br_3^-})}+2c_{Br_3^-}+c_{H^+}\right)\right]=\frac{\partial}{\partial y}\left[p\frac{\partial c_{Br_3^-}}{\partial y}\left(\frac{2D_{Br_2}c_{H^+}}{K(c_{H^+}-c_{Br_3^-})^2}+3D_{Br_3^-}-D_{Br^-}\right)\right].\qquad[1]$$

where $c_i$ and $D_i$ are the concentration and diffusivity of species $i$; $U$ is the local superficial velocity, $p$ the local porosity, and $K$ the complexation reaction equilibrium constant.

The flow in the open electrolyte channel is modelled as Pouseille flow between two flat plates with an area average velocity $U$ between 1.4 and 2.8 cm/s, and the flow in the porous cathode as plug (Darcy) flow at 2.5 cm/s to match the experiments. The porosity, $p$, is set to unity in the open channel, 0.5 in the dispersion blocker, and 0.9 in the porous cathode. The hydronium ion ($H^+$) concentration is fixed throughout the domain at 3 M. The upstream boundary condition is a fixed species concentration, and diffusive flux is set to zero at the downstream boundary and along the surface of the anode and the cathode current collector. Normal fluxes are continuous across all internal boundaries, notably between the dispersion blocker and electrolyte channel.

Species transport in flow-through porous media can be significantly faster than molecular diffusion due to hydrodynamic dispersion, if the flow is fast enough (if Peclet number based on pore radius, $Pe_r = Ur/D \gg 1$).[22] In that case, a large, anisotropic dispersion coefficient would replace the smaller, nearly isotropic molecular diffusion coefficient in the governing species transport equation.[23] In our battery, dispersion would be a key source of oxidant loss from the large-pore, flow-through cathode, were it not for the dispersion blocker in the cathode assembly. This small-pore layer (inside which $Pe_r \ll 1$) prevents dispersive mixing of the oxidant and electrolyte streams and confines dispersion effects to cathode bulk. We thus use molecular diffusivities in Equation 1, corrected by a tortuosity factor within the dispersion blocker, which was a fitting parameter to the measured data (see Results section).

Equation 1 was solved numerically using COMSOL 4.3 software, and the results shown in Figure 1b and 1c. Concentration boundary layers are formed within the porous cathode and electrolyte channel in the vicinity of the dispersion blocker ($Br_3^-$ boundary layers shown in Figure 1b). As shown in Figure 1b, the predicted concentration gradient of $Br_3^-$ is largest within the dispersion blocker by several orders of magnitude. The latter observation indicates that, at these conditions, the dispersion blocker serves as a significantly higher resistor to reactant diffusion than the adjacent concentration boundary layers in the electrolyte channel and cathode. This is in contrast to typical membraneless designs, where the concentration boundary layers between co-flowing laminar streams serve as the sole diffusion resistance between flows.[7,11,13] We note that Hollinger et al. previously observed experimentally that nanoporous separators can inhibit diffusive transport between co-flowing liquid streams,[17] which supports our model conclusions. Thus, our dispersion blocker layer adds an important design knob, as it functions as a large, easily tuneable diffusion resistance. To quantify crossover, in Figure 1c we plot the predicted tribromide retention, defined as the flux of tribromide leaving the downstream edge of the cathode divided by that which enters the upstream edge. We here show that the retention can reach 98.6% for model parameters which mirror those of our proof-of-concept, unoptimized experimental system (see Materials and Methods section).

**Materials and Methods**

The components used in our prototype cell were two PVDF porting plates (McMaster-Carr, Il), a graphite cathode current collector (McMaster-Carr), a composite anode current collector to prevent hydrogen gas leaks (GraphiteStore.com, Il, USA), PTFE coated glass fiber gaskets (American Durafilm, MA) and a fuel cell anode with 0.5 mg/cm$^2$ platinum loading in the catalyst layer supported by a carbon cloth substrate (FuelCellStore.com). The heterogeneous pore structure cathode assembly consisted of a layer of SGL 25AA porous carbon paper as cathode material (SGL Group, Germany), and a layer of porous polyethylene with ~80 nm diameter pores as dispersion blocker (K1650, Celgard, NC). The carbon paper was pre-treated using a 5 h soak at 50°C in a 3:1 volume ratio of 98% $H_2SO_4$ to 70% $HNO_3$ to reduce activation overpotentials.[5] The electrolyte and oxidant channels were cut into the gaskets, and were 1.1 cm long (in the flow direction), 2 mm wide, and 0.6 mm high (in the electric field direction), and this length and width defined the active area of our cell (0.22 cm$^2$ active area).

Polarization curve experiments were performed at room temperature using fresh electrolyte, oxidant, and hydrogen gas. Gas flow was controlled via a mass flow controller (Cole Parmer, Il) set to 25 sccm, and liquid flowrates were controlled via peristaltic pump (Harvard Apparatus, MA) at 1 mL/min through the electrolyte channel and 0.6 mL/min through the cathode. The electrolyte was 3 M HBr, and for discharge curves the oxidant 0.25-3 M $Br_2$ in 3 M HBr (we identify oxidant solutions via their initial, pre-complexation, concentration of $Br_2$). A potentiostat (Reference 3000, Gamry, PA) was used to set the discharge and charge currents and measure the cell voltage. The cell was run at a specified current for 30 s with a voltage measurement occurring at 1 Hz frequency, and then all voltage measurements for a given current were averaged to obtain the polarization curve voltage (the voltage typically took < 5 s to reach steady state). Closed-loop cycling data was obtained by cycling a fixed volume of 10 mL oxidant (0.1 M $Br_2$ in 3 M HBr) and 66.6 mL electrolyte (3 M Hbr) using an electrolyte flow rate of 2 mL/min and oxidant flow rate of 0.6 mL/min, and a constant current of 0.2 A/cm$^2$ during charging and discharging steps. Cell voltage was measured every 5 s, and the half-cycle time was set to 16.7 min to allow a single pass of the oxidant through the system each half-cycle. 0.1 M $Br_2$ was used in order to reduce cycle time needed to attain significant fuel utilization in a single pass, and we here achieve ~26% utilization of the initial $Br_2$ upon discharge.

**Experimental results**

To validate the model presented in the Theory section, we directly measured the tribromide concentration in the electrolyte outflow of our prototype battery using UV spectrophotometry. The spectrophotometry data was obtained using a Shimadzu UV spectrophotometer (Japan), and absorbance measurements were made at a wavelength of 270 nm. A calibration curve was developed of measured absorbance versus known $Br_3^-$ concentration in 3 M HBr solution, and electrolyte samples from the battery were compared to this calibration curve to obtain their $Br_3^-$ concentration. Experiments were run with a 1 M $Br_2$ oxidant stream, and without any current sourced or applied to the cell. As can be seen in Figure 1c, the experimental results were well-described by the model, where for the model we used dispersion blocker tortuosity as a fitting parameter with a value of 1.5 (near to the values measured previously for polypropylene sheets from Celgard[24]).

While our membraneless cell architecture was designed to minimize crossover, it is important that it can also deliver high performance. In Figure 2a, we show the prototype cell's polarization curve data. For charging data (black squares), solely HBr electrolyte was pumped through both the electrolyte channel and porous cathode. Discharging polarization curves were obtained for three concentrations of $Br_2$ (triangular markers). The linearity of the curves at low currents shows that there were no significant activation overpotential losses. At 0.25 M, we can observe some concentration polarization (nonlinearity) above 0.7 A/cm$^2$, but, remarkably, there were no such mass transfer losses for higher bromine concentrations, even at very large currents (3 A/cm$^2$). This suggests that in the future, higher currents are achievable by optimizing the cell for minimum Ohmic resistance. Figure 2a also shows the discharge power density achieved by the prototype (circular markers), calculated from the polarization curve data. Maximum power increased with increasing $Br_2$ concentration in the oxidant flow, up to a value of 0.925 W/cm$^2$ at 3 M. In contrast to the membraneless $H_2$-$Br_2$ flow battery developed by Braff et al. with liquid flow over a planar (flat) electrode,[11] our porous, flow-through architecture exhibits significantly reduced mass transport losses and requires much less bromine to maintain high power (here 0.45 W/cm$^2$ at 0.25 M and 0.79 W/cm$^2$ at 1 M). In addition, our battery achieves a maximum current density of up to 3 A/cm$^2$ (at 3 M $Br_2$), which is nearly twice as high as was previously attained by a membraneless cell (1.7 A/cm$^2$).[11] Further, our prototype demonstrates exceptionally high voltage efficiency, defined as the discharge voltage divided by the charge voltage at a given current, such as 96% efficiency at 3 M $Br_2$ and 20% maximum power (0.19 W/cm$^2$, 0.2 A/cm$^2$).

To put our work in a historical perspective, in Figure 2b we show a review of the maximum power achieved by membraneless electrochemical energy systems (fuel cells and flow batteries).[9–14,17–19,25–45] We distinguish between fuel cells and flow batteries by denoting flow batteries as devices which demonstrated closed-loop charge-discharge cycling. In the inset, we show example closed-loop cycling data obtained from our prototype by plotting the time-averaged charge and discharge cell voltage versus cycle number. To our knowledge, our battery achieves the highest power reported for membraneless electrochemical energy systems (fuel cell or flow battery), and attains roughly 50% higher power density than high performance membrane-based vanadium redox flow batteries (~ 0.6 W/cm$^2$).[46]

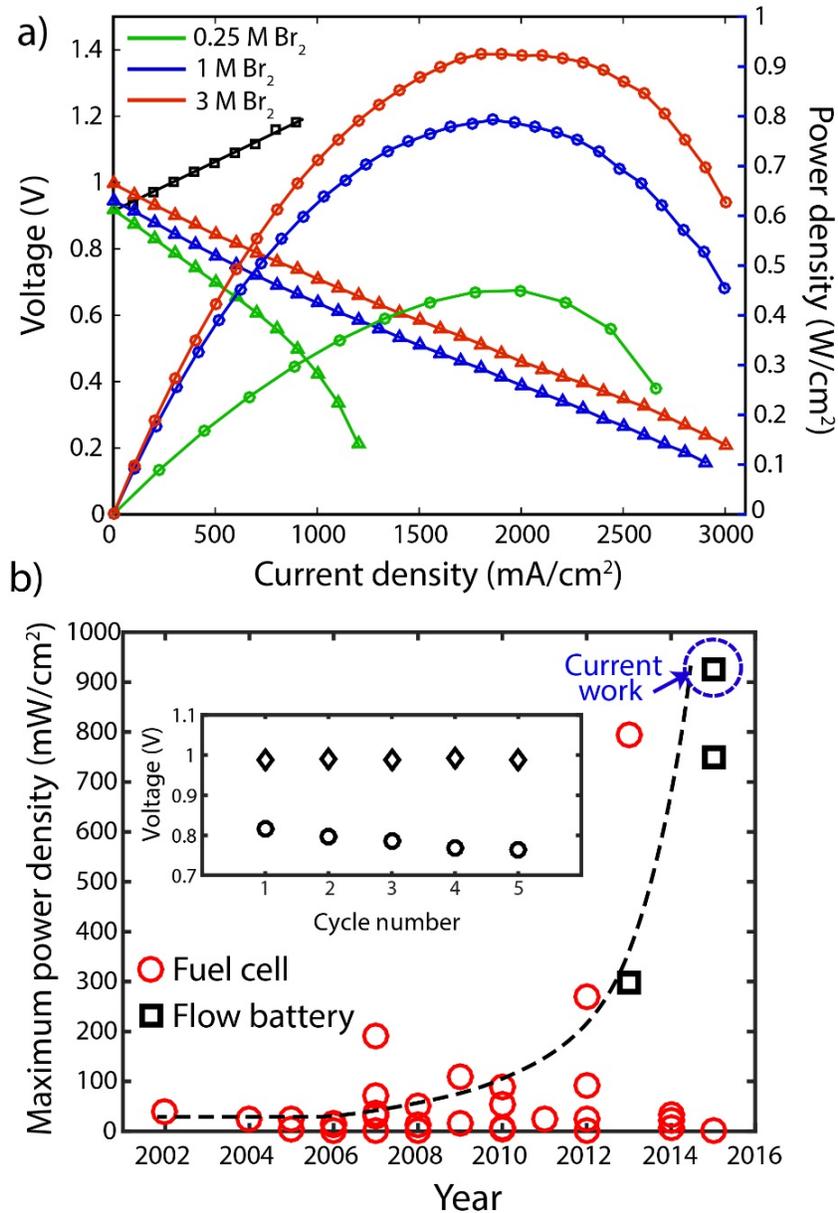

**Figure 2:** a) Measured discharge (triangular markers) and charge (square black markers) polarization curve data, and the corresponding discharge power density (circular markers) versus current density. Our prototype demonstrates a maximum power density of 0.925 W/cm² at 3 M Br₂. b) Historical evolution of the maximum power density attained by membraneless flow batteries and fuel cells. The dashed line serves to guide the eye, and the inset shows the cycling data obtained with our prototype (square markers represent the time-averaged discharge voltage and diamond makers the charge voltage).

## Conclusions

We proposed, numerically modelled, and experimentally demonstrated a novel membraneless flow battery architecture which leverages inexpensive heterogeneous porous media to minimize reactant crossover. In addition, our cell greatly improved upon the state of the art in membraneless flow batteries by attaining a power density of 0.925 W/cm$^2$ and current density of 3 A/cm$^2$. The concept of using heterogeneous flow-through porous media to separate reactants can in the future be applied to a wide variety of redox chemistries.

## Acknowledgements

We would like to acknowledge seed funding from the MIT Energy Initiative (MITEI), a Catalyst Award from the Massachusetts Clean Energy Center (MassCEC), and funding from the Kuwait-MIT Center for Natural Resources and the Environment.